\documentclass[onecolumn,showpacs,preprintnumbers]{revtex4}
\usepackage{graphicx}
\usepackage{dcolumn}
\usepackage{bm}
\usepackage{epsfig}
\usepackage{amsmath,amssymb}

\begin{document}
\title{Systematic analysis of the $D_{J}(2580)$, $D_{J}^{*}(2650)$, $D_{J}(2740)$, $D_{J}^{*}(2760)$, $D_{J}(3000)$ and $D_{J}^{*}(3000)$ in $D$ meson family
}
\author{Guo-Liang Yu$^{1}$}
\email{yuguoliang2011@163.com}
\author{Zhi-Gang Wang$^{1}$}
\email{zgwang@aliyun.com}
\author{Zhen-Yu Li$^{2}$}

\affiliation{$^1$ Department of Mathematics and Physics, North China
Electric power university, Baoding 071003, People's Republic of China\\$^2$
School of Physics and Electronic Science, Guizhou Normal College,
Guiyang 550018, People's Republic of China}
\date{\today }

\begin{abstract}
In this work, we tentatively assign the charmed mesons
$D_{J}(2580)$, $D_{J}^{*}(2650)$, $D_{J}(2740)$, $D_{J}^{*}(2760)$,
$D_{J}(3000)$ and $D_{J}^{*}(3000)$ observed by the LHCb
collaboration according to their spin, parity and masses, then
systematically study their strong decays to the ground state charmed
mesons plus pseudoscalar mesons with the $^{3}P_{0}$ decay model.
According to these studies, we assign the $D_{J}^{*}(2760)$ as the
$1D\frac{5}{2}3^{-}$ state, the $D_{J}^{*}(3000)$ as the
$1F\frac{5}{2}2^{+}$ or $1F\frac{7}{2}4^{+}$ state, the
$D_{J}(3000)$ as the $1F\frac{7}{2}3^{+}$ or $2P\frac{1}{2}1^{+}$
state in the $D$ meson family. As a byproduct, we also study the
strong decays of $2P\frac{1}{2}0^{+}$,$2P\frac{3}{2}2^{+}$,
$3S\frac{1}{2}1^{-}$, $3S\frac{1}{2}0^{-}$ etc, states, which will
be valuable in
 searching for the partners of these $D$ mesons.
\end{abstract}


\maketitle

\begin{large}
\textbf{1 Introduction}
\end{large}

In 2013, the LHCb Collaboration announced several $D_{J}$ resonances
by studying the $D^{+}\pi^{-}$, $D^{0}\pi^{+}$, $D^{*+}\pi^{-}$
invariant mass spectra, which were obtained from the $pp$ collisions
at a center-of-mass energy of 7TeV~\cite{Aaij}. The LHCb
Collaboration observed two natural parity resonances
$D_{J}^{*}(2650)^{0}$, $D_{J}^{*}(2760)^{0}$  and two unnatural
parity resonances $D_{J}(2580)^{0}$, $D_{J}(2740)^{0}$ in the
$D^{*+}\pi^{-}$ mass spectrum, and tentatively identified
$D_{J}(2580)$ as the $2S$ $0^{-}$ state, the $D_{J}^{*}(2650)$ as
the $2S$ $1^{-}$ state, the $D_{J}(2740)$ as the $1D$ $2^{-}$ state,
the $D_{J}^{*}(2760)$ as the $1D$ $1^{-}$ state, respectively. The
$D_{J}^{*}(2760)^{0}$ observed in the $D^{*+}\pi^{-}$ and
$D^{+}\pi^{-}$ decay modes have consistent parameters, their charged
partner $D_{J}^{*}(2760)^{+}$ was observed in the $D^{0}\pi^{+}$
final state~\cite{Aaij}. Furthermore, the LHCb collaboration also
observed one unnatural parity resonance $D_{J}(3000)^{0}$ in the
$D^{*+}\pi^{-}$ final state, and two resonances
$D_{J}^{*}(3000)^{0}$ and $D_{J}^{*}(3000)^{+}$ in the
$D^{+}\pi^{-}$ and $D^{0}\pi^{+}$ mass spectra,
respectively~\cite{Aaij}. The relevant parameters are presented in
Table I.

\begin{table*}[t]
\begin{ruledtabular}\caption{The experimental results from the LHCb collaboration, where the N and U denote the
natural parity and unnatural parity, respectively.}
\begin{tabular}{c c c c c c c c c c c c c c c c c c}
   & \ Mass(MeV)  & \ Width(MeV)  & \ Decay Channel & \ Significance   \\
\hline
$D_{J}^{*}(2650)^{0}$   (N)    &  \  $2649.2\pm3.5\pm3.5$      & \  $140.2\pm17.1\pm18.6$       &  \   $D^{*+}\pi^{-}$   & \  $24.5\sigma$      \\
$D_{J}^{*}(2760)^{0}$   (N)    &  \  $2761.1\pm5.1\pm6.5$      & \  $74.4\pm3.4\pm37.0$       &  \   $D^{*+}\pi^{-}$   & \  $10.2\sigma$      \\
$D_{J}(2580)^{0}$   (U)    &  \  $2579.5\pm3.4\pm5.5$      & \  $177.5\pm17.8\pm46.0$       &  \   $D^{*+}\pi^{-}$   & \  $18.8\sigma$      \\
$D_{J}(2740)^{0}$   (U)    &  \  $2737.0\pm3.5\pm11.2$      & \  $73.2\pm13.4\pm25.0$       &  \   $D^{*+}\pi^{-}$   & \  $7.2\sigma$      \\
$D_{J}(3000)^{0}$   (U)    &  \  $2971.8\pm8.7$      & \  $188.1\pm44.8$       &  \   $D^{*+}\pi^{-}$   & \  $9.0\sigma$      \\
\hline
$D_{J}^{*}(2760)^{0}$   (N)    &  \  $2760.1\pm1.1\pm3.7$      & \  $74.4\pm3.4\pm19.1$       &  \   $D^{+}\pi^{-}$   & \  $17.3\sigma$      \\
$D_{J}^{*}(3000)^{0}$          &  \  $3008.1\pm4.0$      & \  $110.5\pm11.5$       &  \   $D^{+}\pi^{-}$   & \  $21.2\sigma$      \\
\hline
$D_{J}^{*}(2760)^{+}$   (N)    &  \  $2771.7\pm1.7\pm3.8$      & \  $66.7\pm6.6\pm10.5$       &  \   $D^{0}\pi^{+}$   & \  $18.8\sigma$      \\
$D_{J}^{*}(3000)^{+}$   (N)    &  \  $3008.1$ (fixed)      & \  $110.5$ (fixed)       &  \   $D^{0}\pi^{+}$   & \  $6.6\sigma$      \\
\end{tabular}
\end{ruledtabular}
\end{table*}

\begin{table*}[t]
\begin{ruledtabular}\caption{The experimental results from the BaBar collaboration, the particles in the bracket are
the possible corresponding ones observed by the LHCb collaboration.}
\begin{tabular}{c c c c c c c c c c c c c c c c c c}
   & \ Mass(MeV)  & \ Width(MeV)  & \ Decay Channel   \\
\hline
$D^{0}(2550)$ [$D_{J}(2580)^{0}$]   &  \  $2539.4\pm4.5\pm6.8$      & \  $130\pm12\pm13$       &  \   $D^{*+}\pi^{-}$        \\
$D^{0}(2600)$ [$D_{J}^{*}(2650)^{0}$]   &  \  $2608.7\pm2.4\pm2.5$      & \  $93\pm6\pm13$       &  \   $D^{+}\pi^{-}$,$D^{*+}\pi^{-}$        \\
$D^{0}(2750)$ [$D_{J}(2740)^{0}$]   &  \  $2752.4\pm1.7\pm2.7$      & \  $71\pm6\pm11$       &  \   $D^{*+}\pi^{-}$        \\
$D^{0}(2760)$ [$D_{J}^{*}(2760)^{0}$]   &  \  $2763.3\pm2.3\pm2.3$      & \  $60.9\pm5.1\pm3.6$       &  \   $D^{+}\pi^{-}$        \\
$D^{+}(2600)$    &  \  $2621.3\pm3.7\pm4.2$      & \  $93$       &  \   $D^{0}\pi^{+}$        \\
$D^{+}(2760)$ [$D_{J}^{*}(2760)^{+}$]   &  \  $2769.7\pm3.8\pm1.5$      & \  $60.9$       &  \   $D^{0}\pi^{+}$        \\
\end{tabular}
\end{ruledtabular}
\end{table*}

In 2010, the BaBar collaboration observed four excited charmed
mesons $D(2550)$, $D(2600)$, $D(2750)$ and $D(2760)$ in the decays
$D^{0}(2550)\rightarrow D^{*+}\pi^{-}$, $D^{0}(2600)\rightarrow
D^{*+}\pi^{-}$, $D^{+}\pi^{-}$, $D^{0}(2750)\rightarrow
D^{*+}\pi^{-}$, $D^{0}(2760)\rightarrow D^{+}\pi^{-}$,
$D^{+}(2600)\rightarrow D^{0}\pi^{+}$ and $D^{+}(2760)\rightarrow
D^{0}\pi^{+}$ respectively in the inclusive $e^{+}e^{-}\rightarrow
c\overline{c}$ interactions~\cite{Amo}. The BaBar collaboration also
analyzed the helicity distributions to determine the spin-parity,
and tentatively identified the ($D(2550)$,$D(2600)$) as the $2 S$
doublet ($0^{-}, 1^{-}$), the $D(2750)$ and $D(2760)$ as the D-wave
states. The relevant  parameters are presented in Table II, where we
also present the possible correspondences among the particles
observed by the LHCb and BaBar collaborations. The physicists have
also studied the decay behaviors of these charmed mesons using the
heavy meson effective theory~\cite{Wang11}, constituent quark
model~\cite{Zhong10} and the Eichten-Hill-Quigg's
formula~\cite{Chen11}.

The heavy meson effective theory is a powerful tool in studying  the
properties of hadrons with a single heavy quark. With this method,
P. Colangelo \emph{et al}. proposed a classification of many
observed $c\overline{q}$ and $b\overline{q}$ mesons in
doublets~\cite{Cola12}. In Ref.~\cite{Wang}, we study the strong
decays of the charmed mesons $D_{J}(2580)$, $D_{J}^{*}(2650)$,
$D_{J}(2740)$, $D_{J}^{*}(2760)$, $D_{J}(3000)$ and
$D_{J}^{*}(3000)$  with the heavy meson effective theory in the
leading order approximation. And the ratios among decay widths of
different channels were calculated. But the exact value of the decay
widths were not given out, which constitutes the first motivation of
our study. The quark pair creation (QPC) model is another effective
method to study the strong decays of the mesons, which is also known
as the $^{3}P_{0}$ decay model. It was originally introduced by L.
Micu~\cite{Micu} and further developed by A. Le Yaouanc \emph{et
al}.~\cite{Le}. This model has been widely used to evaluate the
strong decays of
hadrons~\cite{Yao77,Yaou88,Robe92,Capstick,Blun,Geiger,Zhou05,Chen09,Dml08,Dml10,Bzh07,Yang10},
since it gives a good description of many  observed decay amplitudes
and partial widths of the hadrons. Y. Sun \emph{et al}.~\cite{Sun13}
studied the strong decays of the $D_{J}(3000)$ and $D_{J}^{*}(3000)$
with the $^{3}P_{0}$ decay model, and identified $D_{J}(3000)$ as
the $2P(1^{+})$ state, the $D_{J}^{*}(3000)$ as the $2^{3}P_{0}$
state, respectively. But $D_{J}(2580)$, $D_{J}^{*}(2650)$,
$D_{J}(2740)$ and $D_{J}^{*}(2760)$, which were also observed by
LHCb Collaboration, were not analyzed in their studies. This is the
second motivation of our work. Besides, they chose the simple
harmonic oscillator (SHO) wave functions with the effective
oscillator parameter $R$ as the meson's radial wave functions. From
Ref.~\cite{Blun}, we can see that there are two types of SHO wave
functions: SHO wave functions with a common oscillator parameter $R$
and with an  effective oscillator parameter $R$. According to a
series of least squares fits of the model predictions to the decay
widths of 28 of the best known meson decays, it seems  that the  SHO
wave functions with a common $R$ can lead to better
results~\cite{Blun}. Thus, in order to identify the $D_{J}(2580)$,
$D_{J}^{*}(2650)$, $D_{J}(2740)$, $D_{J}^{*}(2760)$, $D_{J}(3000)$
and $D_{J}^{*}(3000)$, it is necessary and interesting to
systematically study the strong decays of these charmed mesons  by
the $^{3}P_{0}$ decay model with the common oscillator parameter
$R$.

In the heavy quark limit, the  heavy-light mesons $Q\overline{q}$
can be classified in doublets according to the total angular
momentum of the light antiquark $\vec s_{l}$, $\vec s_{l}$= $\vec
s_{\overline{q}}$+$\vec L$, where the $\vec s_{\overline{q}}$ and
$\vec {L}$ are the spin and orbital angular momentum of the light
antiquark, respectively~\cite{Mano}. In the case of the radial
quantum number $n = 1$, the doublet ($P$,$P^{*}$) have the
spin-parity $J_{s_{l}}^{P}=(0^{-},1^{-})_{\frac{1}{2}}$ for $L = 0$;
the two doublets ($P_{0}^{*}$, $P_{1}$) and ($P_{1}$, $P_{2}^{*}$)
have the spin-parity $J_{s_{l}}^{P}=(0^{+},1^{+})_{\frac{1}{2}}$ and
$(1^{+},2^{+})_{\frac{3}{2}}$ respectively for $L = 1$; the two
doublets ($P_{1}^{*}$, $P_{2}$) and ($P_{2}$, $P_{3}^{*}$) have the
spin-parity $J_{s_{l}}^{P}=(1^{-},2^{-})_{\frac{3}{2}}$ and
$(2^{-},3^{-})_{\frac{5}{2}}$ respectively for $L = 2$; the two
doublets ($P_{2}^{*}$, $P_{3}$) and ($P_{3}$, $P_{4}^{*}$) have the
spin-parity $J_{s_{l}}^{P}=(2^{+},3^{+})_{\frac{5}{2}}$ and
$(3^{+},4^{+})_{\frac{7}{2}}$ respectively for $L = 3$, where the
superscript $P$ denotes the parity. The $n = 2, 3, 4, ¡¤ ¡¤ ¡¤$
states are clarified by analogous doublets, for example, $n = 2$,
the doublet ($P^{'}, P^{*'}$) have the spin-parity $J_{s_{l}}^{P}$ =
$(0^{-},1^{-})_{\frac{1}{2}}$ for $L = 0$.

The $D_{J }(2580)^{0}$, $D_{J} (2740)^{0}$ and $D_{J} (3000)^{0}$
have unnatural parity, and their possible spin-parity assignments
are $J^{P}$ = $0^{-}$, $1^{+}$, $2^{-}$, $3^{+}$, $\cdots$. The
$D_{J}^{*} (2650)^{0}$ and $D_{J}^{*} (2760)^{0}$ and $D_{J}^{*}
(3000)$ have natural parity, and their possible spin-parity
assignments are $J^{P}$ = $0^{+}$, $1^{-}$, $2^{+}$, $3^{-}$,
$\cdots$. The six low-lying states, $D$, $D^{*}$, $D_{0}(2400)$,
$D_{1}(2430)$, $D_{1}(2420)$ and $D_{2}(2460)$ have been
established~\cite{Beri}. The newly observed charmed mesons $D_{J}
(2580)$, $D_{J}^{*} (2650)$, $D_{J} (2740)$, $D_{J}^{*} (2760)$,
$D_{J} (3000)$, $D_{J}^{*} (3000)$ can be tentatively identified as
the missing states in the $D$ meson family.

The mass is a fundamental parameter in describing a hadron, in TABLE
III, we present the predictions from some theoretical models, such
as the relativized quark model based on a universal one-gluon
exchange plus linear confinement potential~\cite{Godf}, the
relativistic quark model includes the leading order $1/M_{h}$
corrections~\cite{Pierro}, the QCD-motivated relativistic quark
model based on the quasipotential approach~\cite{Eber}. We can
identify the $D_{J} (2580)$, $D_{J}^{*} (2650)$, $D_{J} (2740)$,
$D_{J}^{*} (2760)$, $D_{J} (3000)$, $D_{J}^{*} (3000)$ tentatively
according to the masses.

\begin{table*}[htbp]
\begin{ruledtabular}\caption{The masses of the charmed mesons from different quark models compared with experimental data,
and the possible assignments of the newly observed charmed mesons. The N and U
denote the natural parity and unnatural parity, respectively. All
values in units of MeV.}
\begin{tabular}{c c c c c c c c c c c c c c c c c c}
       & \ $n L s_{L} J^{P}$  & \ Exp~\cite{Aaij,Beri}  & \ GI~\cite{Aaij,Godf} & \ PE~\cite{Pierro} & \ EFG~\cite{Eber}   \\
\hline
 $D$              &  \  $1 S \frac{1}{2} 0^{-}$    & \  $1867$    &  \   $1864$   & \  $1868$     &  \  $1871$        \\
 $D^{*}$          &  \  $1 S \frac{1}{2} 1^{-}$    & \  $2008$    &  \   $2023$   & \  $2005$     &  \  $2010$        \\
\hline
 $D^{*}_{0}$      &  \  $1 P \frac{1}{2} 0^{+}$    & \  $2400$    &  \   $2380$   & \  $2377$     &  \  $2406$        \\
 $D_{1}$          &  \  $1 P \frac{1}{2} 1^{+}$    & \  $2427$    &  \   $2419$   & \  $2490$     &  \  $2469$        \\
 $D_{1}$          &  \  $1 P \frac{3}{2} 1^{+}$    & \  $2420$    &  \   $2469$   & \  $2417$     &  \  $2426$        \\
 $D^{*}_{2}$      &  \  $1 P \frac{3}{2} 2^{+}$    & \  $2460$    &  \   $2479$   & \  $2460$     &  \  $2460$        \\
\hline
 $D^{*}_{1}$      &  \  $1 D \frac{3}{2} 1^{-}$    & \  ?$2760$(N)    &  \   $2796$   & \  $2795$     &  \  $2788$        \\
 $D_{2}$          &  \  $1 D \frac{3}{2} 2^{-}$    & \  ?$2740$(U)    &  \   $2801$   & \  $2833$     &  \  $2850$        \\
 $D_{2}$          &  \  $1 D \frac{5}{2} 2^{-}$    & \  ?$2740$(U)    &  \   $2806$   & \  $2775$     &  \  $2806$        \\
 $D^{*}_{3}$      &  \  $1 D \frac{5}{2} 3^{-}$    & \  ?$2760$(N)    &  \   $2806$   & \  $2799$     &  \  $2863$        \\
\hline
 $D^{*}_{2}$      &  \  $1 F \frac{5}{2} 2^{+}$    & \  ?$3000$(N)    &  \   $3074$   & \  $3101$     &  \  $3090$        \\
 $D_{3}$          &  \  $1 F \frac{5}{2} 3^{+}$    & \  ?$3000$(U)    &  \   $3074$   & \  $3123$     &  \  $3145$        \\
 $D_{3}$          &  \  $1 F \frac{7}{2} 3^{+}$    & \  ?$3000$(U)    &  \   $3079$   & \  $3074$     &  \  $3129$        \\
 $D^{*}_{4}$      &  \  $1 F \frac{7}{2} 4^{+}$    & \  ?$3000$(N)    &  \   $3084$   & \  $3091$     &  \  $3187$        \\
\hline
 $D$              &  \  $2 S \frac{1}{2} 0^{-}$    & \  ?$2580$(U)    &  \   $2558$   & \  $2589$     &  \  $2581$        \\
 $D^{*}$          &  \  $2 S \frac{1}{2} 1^{-}$    & \  ?$2650$(N)    &  \   $2618$   & \  $2692$     &  \  $2632$        \\
\hline
 $D^{*}_{0}$      &  \  $2 P \frac{1}{2} 0^{+}$    & \  ?$3000$(N)    &  \            & \  $2949$     &  \  $2919$        \\
 $D_{1}$          &  \  $2 P \frac{1}{2} 1^{+}$    & \  ?$3000$(U)    &  \            & \  $3045$     &  \  $3021$        \\
 $D_{1}$          &  \  $2 P \frac{3}{2} 1^{+}$    & \  ?$3000$(U)    &  \            & \  $2995$     &  \  $2932$        \\
 $D^{*}_{2}$      &  \  $2 P \frac{3}{2} 2^{+}$    & \  ?$3000$(N)    &  \            & \  $3035$     &  \  $3012$        \\
\hline
 $D$              &  \  $3 S \frac{1}{2} 0^{-}$    & \  ?$3000$(U)    &  \            & \  $3141$     &  \  $3062$        \\
 $D^{*}$          &  \  $3 S \frac{1}{2} 1^{-}$    & \  ?$3000$(N)    &  \            & \  $3226$     &  \  $3096$        \\
\end{tabular}
\end{ruledtabular}
\end{table*}

In the following, we list out the possible assignments,
\begin{equation*}
(D_{J} (2580),D_{J}^{*}(2650)) = (0^{-}, 1^{-})_{\frac{1}{2}} \quad  with \quad   n = 2, L = 0\, ,
\end{equation*}
\begin{equation*}
(D_{J}^{*} (2760),D_{J}(2740)) = (1^{-}, 2^{-})_{\frac{3}{2}}\quad  with \quad n = 1, L = 2\, ,
\end{equation*}
\begin{equation*}
(D_{J} (2740),D_{J}^{*}(2760)) = (2^{-}, 3^{-})_{\frac{5}{2}}\quad
with \quad n = 1, L = 2 \, ,
\end{equation*}
\begin{equation*}
(D_{J}^{*} (3000),D_{J}(3000)) = (2^{+}, 3^{+})_{\frac{5}{2}}\quad
with \quad n = 1, L = 3 \, ,
\end{equation*}
\begin{equation*}
(D_{J} (3000),D_{J}^{*}(3000)) = (3^{+}, 4^{+})_{\frac{7}{2}}\quad
with \quad n = 1, L = 3 \, ,
\end{equation*}
\begin{equation*}
(D_{J}^{*} (3000),D_{J}(3000)) = (0^{+}, 1^{+})_{\frac{1}{2}}\quad
with \quad n = 2, L = 1 \, ,
\end{equation*}
\begin{equation*}
(D_{J} (3000),D_{J}^{*}(3000)) = (1^{+}, 2^{+})_{\frac{3}{2}}\quad
with \quad n = 2, L = 1\, ,
\end{equation*}
\begin{equation*}
(D_{J} (3000),D_{J}^{*}(3000)) = (0^{-}, 1^{-})_{\frac{1}{2}}\quad
with \quad n = 3, L = 0\, .
\end{equation*}

The article is arranged as follows: In section 2, the brief review
of the $^{3}P_{0}$ decay model is given (For the detailed review see
Refs.~\cite{Le,Yaou88,Robe92,Blun}); In section 3, we study the
strong decays of the charmed mesons $D_{J} (2580)$, $D_{J}^{*}
(2650)$, $D_{J} (2740)$, $D_{J}^{*} (2760)$, $D_{J} (3000)$,
$D_{J}^{*} (3000)$ with the $^{3}P_{0}$ decay model; In section 4,
we present our conclusions.

\begin{large}
\textbf{2 METHOD}
\end{large}

\begin{large}
\textbf{2.1 The decay model}
\end{large}

\begin{figure}[h]
  \includegraphics[width=14cm]{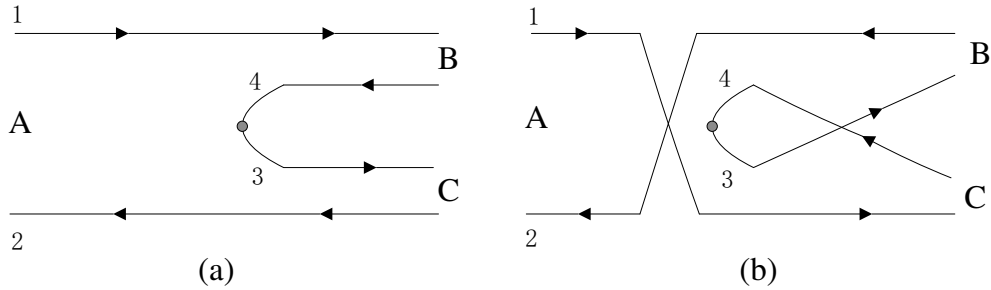}
  \caption{The two possible diagrams contributing to $A\rightarrow B + C$ in the $^{3}P_{0}$ decay  model.}\label{Figure 1:}
\end{figure}

The main assumption of the $^{3}P_{0}$ decay model is that the strong decays take place via the
creation of a $^{3}P_{0}$ quark-antiquark pair from the vacuum. The new produced quark-antiquark
pair, together with the $q\overline{q}$  in the initial meson, regroups into two outgoing mesons in all
possible quark rearrangement ways, which corresponds to the two Feynman  diagrams as shown in
Fig.1 for the strong  decay processes  $A$$\rightarrow$$B + C$.

The transition operator $T$ of the decay $A \rightarrow B+  C$ in the $^{3}P_{0}$ decay  model is given by
\begin{eqnarray}
    T=-3\gamma\sum_{m}\langle1m1-m\mid00\rangle\int d^{3}\vec p_{3}d^{3}\vec p_{4}\delta^{3}(\vec p_{3}+\vec p_{4})
    \mathcal{Y}_{1}^{m}(\frac{\vec p_{3}-\vec p_{4}}{2})\chi_{1-m}^{34}\phi_{0}^{34}\omega_{0}^{34}b_{3}^{\dag}(\vec p_{3})d_{4}^{\dag}(\vec
    p_{4}) \, ,
\end{eqnarray}
where $\gamma$ is a dimensionless parameter representing the
probability of the quark-antiquark pair $q_{3}\overline{q}_{4}$ with
$J^{PC} = 0^{++}$ created  from the vacuum, $\vec p_{3}$ and $\vec
p_{4}$ are the momenta of the created quark $q_{3}$ and antiquark
$\overline{q}_{4}$, respectively. $\phi_{0}^{34}$,
$\omega_{0}^{34}$, and $\chi_{1-m}^{34}$
 are the flavor, color, and spin
wave functions of the $q_{3}\overline{q}_{4}$, respectively. The
solid harmonic polynomial $\mathcal{Y}_{1}^{m}(\vec p)\equiv|\vec
p|^{1}Y_{1}^{m}(\theta_{p},\phi_{p})$ reflects the momentum-space
distribution of the $q_{3}\overline{q}_{4}$.

For the meson wave function, we adopt the mock meson state $ \mid
A(n_{A}^{2S_{A}+1}L_{AJ_{A}M_{J_{A}}})(\vec P_{A})\rangle$ defined
by~\cite{Hayn}
\begin{equation}
\begin{split}
    \mid A(n_{A}^{2S_{A}+1}L_{AJ_{A}M_{J_{A}}})(\vec P_{A})\rangle\equiv
    &\sqrt{2E_{A}}\sum_{M_{L_{A}}M_{S_{A}}}\langle L_{A}M_{L_{A}}S_{A}M_{S_{A}}\mid J_{A}M_{J_{A}}\rangle \\
   &\times \int d^{3}\vec p_{A}\psi_{n_{A}L_{A}M_{L_{A}}}(\vec p_{A})\chi_{S_{A}M_{S_{A}}}^{12}\phi_{A}^{12}\omega_{A}^{12} \\
   &\times \mid q_{1}(\frac{m_{1}}{m_{1}+m_{2}}\vec P_{A}+\vec p_{A})q_{2}(\frac{m_{2}}{m_{1}+m_{2}}\vec P_{A}-\vec p_{A})\rangle \, ,
\end{split}
\end{equation}
where $m_{1}$ and $m_{2}$ are the masses of the quark $q_{1}$ with a
momentum of $\vec p_{1}$ and the antiquark $\overline{q}_{2}$ with a
momentum of $\vec p_{2}$, respectively, $n_{A}$ is the radial
quantum number of the meson $A$ composed of $q_{1}\overline{q}_{2}$,
$\vec S_{A} = \vec s_{q_{1}} + \vec s_{q_{2}}$, $ \vec J_{A} = \vec
L_{A} + \vec S_{A}$, $\vec s_{q_{1}}$($\vec s_{q_{2}}$) is the spin
of $q_{1}$($\overline{q}_{2}$), $\vec L_{A}$ is the relative orbital
angular momentum between $q_{1}$ and $\overline{q}_{2}$, $\vec P_{A}
= \vec p_{1} + \vec p_{2}$, $\vec p_{A} = \frac{m_{2}\vec
p_{1}-m_{1}\vec p_{2}}{m_{1}+m_{2}}$, $\langle
L_{A}M_{L_{A}}S_{A}M_{S_{A}}|J_{A}M_{J_{A}}\rangle$ is a
Clebsch-Gordan coefficient, and $E_{A}$ is the total energy of the
meson $A$, $\chi_{S_{A}M_{S_{A}}}^{12}$, $\phi_{A}^{12}$,
$\omega_{A}^{12}$, and $\psi_{n_{A}L_{A}M_{L_{A}}}(\vec p_{A})$ are
the spin, flavor, color, and space wave functions of the meson $A$,
respectively.

The $S$-matrix of the process $A\rightarrow BC$ is defined by

\begin{equation}
 \langle BC\mid S\mid A\rangle=I-2\pi i\delta(E_{A}-E_{B}-E_{C})\langle BC\mid T\mid
 A\rangle \, ,
\end{equation}
with
\begin{equation}
\langle BC\mid T\mid A\rangle=\delta^{3}(\vec P_{A}-\vec P_{B}-\vec P_{C})\mathcal{M}^{M_{J_{A}}M_{J_{B}}M_{J_{C}}} \, ,
\end{equation}
where $\mathcal{M}^{M_{J_{A}}M_{J_{B}}M_{J_{C}}}$ is the helicity amplitude of $A\rightarrow BC$. In the center of mass frame of
the meson $A$, the $\mathcal{M}^{M_{J_{A}}M_{J_{B}}M_{J_{C}}}$ can be written as

\begin{equation}
\begin{aligned}
\mathcal{M}^{M_{J_{A}}M_{J_{B}}M_{J_{C}}}(\vec P)=
&\gamma\sqrt{8E_{A}E_{B}E_{C}}
\sum_{\mbox{\tiny$\begin{array}{c}
M_{L_{A}},M_{S_{A}},\\
M_{L_{B}},M_{S_{B}},\\
M_{L_{C}},M_{S_{C}},m\end{array}$}}\langle L_{A}M_{L_{A}}S_{A}M_{S_{A}}\mid J_{A}M_{J_{A}}\rangle \\
&\times\langle L_{B}M_{L_{B}}S_{B}M_{S_{B}}\mid J_{B}M_{J_{B}}\rangle
\langle L_{C}M_{L_{C}}S_{C}M_{S_{C}}\mid J_{C}M_{J_{C}}\rangle \\
&\times \langle 1m1-m\mid 00\rangle\langle \chi_{S_{B}M_{S_{B}}}^{14}\chi_{S_{C}M_{S_{C}}}^{32}\mid \chi_{S_{A}M_{S_{A}}}^{12}\chi_{1-m}^{34}\rangle \\
&\times\left[\langle \phi_{B}^{14}\phi_{C}^{32}\mid \phi_{A}^{12}\phi_{0}^{34}\rangle I(\vec P,m_{1},m_{2},m_{3})\right. \\
&\left.+(-1)^{1+S_{A}+S_{B}+S_{C}}\langle \phi_{B}^{32}\phi_{C}^{14}\mid \phi_{A}^{12}\phi_{0}^{34}\rangle I(-\vec P,m_{2},m_{1},m_{3})\right] \, ,
\end{aligned}
\end{equation}
where the two terms in the bracket $[~]$ correspond to the two
possible diagrams in Fig.1(a) and 1(b), respectively, and the
spatial integral is defined as

\begin{equation}
\begin{split}
I(\vec P,m_{1},m_{2},m_{3})=
&\int d^{3}\vec p \psi^{*}_{n_{B}L_{B}M_{L_{B}}}(\frac{m_{3}}{m_{1}+m_{2}}\vec P_{B}+\vec p)\psi^{*}_{n_{C}L_{C}M_{L_{C}}}(\frac{m_{3}}{m_{2}+m_{3}}\vec P_{B}+\vec p) \\
&\times\psi_{n_{A}L_{A}M_{L_{A}}}(\vec P_{B}+\vec p)\mathcal{Y}_{1}^{m}(\vec p) \, ,
\end{split}
\end{equation}
where $\vec P =\vec P_{B} =-\vec P_{C}, \vec p = \vec p_{3}$, $m_{3}$ is the mass of the created quark $q_{3}$, the  SHO
approximation is used for the meson's radial wave functions. In momentum-space, the SHO wave function is
\begin{equation}
\begin{split}
\Psi_{nLM_{L}}(\vec p)=
&(-1)^{n}(-i)^{L}R^{L+\frac{3}{2}}\sqrt{\frac{2n!}{\Gamma(n+L+\frac{3}{2})}} \\
&\times \exp(-\frac{R^{2}p^{2}}{2})L_{n}^{L+\frac{1}{2}}(R^{2}p^{2})\mathcal{Y}_{LM_{L}}(\vec p) \, ,
\end{split}
\end{equation}
here $\mathcal{Y}_{LM_{L}}(\vec p)=|\vec
p|^{L}Y_{LM_{L}}(\Omega_{p})$, and
$L_{n}^{L+\frac{1}{2}}(R^{2}p^{2})$ is an associated Laguerre
polynomial. The overlaps of the flavor and spin wave functions of
the mesons and the created pair in the formula(5) can be calculated
according to the method  in Ref.~\cite{Blun}.

With the Jacob-Wick formula the helicity amplitude can be converted
into the partial wave amplitude~\cite{Jaco}
\begin{equation}
\begin{split}
\mathcal{M}^{JL}(\vec P)=
&\frac{\sqrt{4\pi(2L+1)}}{2J_{A}+1}\sum_{M_{J_{B}}M_{J_{C}}}\langle L0JM_{J_{A}}|J_{A}M_{J_{A}}\rangle \\
&\times\langle J_{B}M_{J_{B}}J_{C}M_{J_{C}}|JM_{J_{A}}\rangle\mathcal{M}^{M_{J_{A}}M_{J_{B}}M_{J_{C}}}(\vec P)\, .
\end{split}
\end{equation}
The decay width in terms of the partial wave amplitudes using the
relativistic phase space is

\begin{equation}
\Gamma=\frac{\pi}{4}\frac{|\vec P|}{M_{A}^{2}}\sum_{JL}|\mathcal{M}^{JL}|^{2}\, ,
\end{equation}
where $|\vec P|=\frac{\sqrt{[M_{A}^{2}-(M_{B}+M_{C})^{2}][M_{A}^{2}-(M_{B}-M_{C})^{2}]}}{2M_{A}}$, $M_{A}$, $M_{B}$, and $M_{C}$ are the masses of the mesons
$A$, $B$, and $C$, respectively.

\begin{large}
\textbf{2.2 Mixed states}
\end{large}

The heavy-light mesons are not charge conjugation eigenstates and so
mixing can occur between states with $J=L$ and $S=1$ or $0$. A
general relation between the heavy quark symmetric states and the
non-relativistic states $^{3}L_{L}$ and $^{1}L_{L}$ can be written
as~\cite{Mats}

\begin{equation}
\begin{pmatrix} |s_{l}=L-\frac{1}{2},L^{P}\rangle &  \\ |s_{l}=L+\frac{1}{2},L^{P}\rangle  \end{pmatrix} \\=\frac{1}{\sqrt{2L+1}}\begin{pmatrix} \sqrt{L+1} &  -\sqrt{L}  \\ \sqrt{L}&  \sqrt{L+1}   \end{pmatrix} \\\begin{pmatrix} |^{3}L_{L}\rangle &  \\ |^{1}L_{L}\rangle  \end{pmatrix} \\ \, , P=(-1)^{L+1} \, .
\end{equation}
Commonly, we express this relation with the mixture. When $J=L=1$,
the corresponding mixture angle is $\theta=-54.7^{\circ}$ or
$\theta=35.3^{\circ}$, thus formula (10) transforms into

\begin{equation}
\begin{pmatrix} |\frac{1}{2},1^{+}\rangle &  \\ |\frac{3}{2},1^{+}\rangle  \end{pmatrix} \\=\begin{pmatrix} \cos\theta &  -\sin\theta  \\ \sin\theta & \cos\theta   \end{pmatrix} \\\begin{pmatrix} |^{3}P_{1}\rangle &  \\ |^{1}P_{1}\rangle   \end{pmatrix} \, .\\
\end{equation}
In our calculation, the final states are related to
$D(2420)/D(2430)$ and $D_{s}(2460)/D_{s}(2536)$, which are the
$1^{+}$ states in the $D$ and $D_{s}$ meson families, respectively.
The $D(2420)/D(2430)$ and $D_{s}(2460)/D_{s}(2536)$ are the mixings
of the $^{3}P_{1}$ and $^{1}P_{1}$ states, which satisfies the
formula (11). In addition, the initial states of $1^{+}$ are also
the mixings of $^{3}P_{1}$ and $^{1}P_{1}$ states. As far as the
$1F\frac{5}{2}3^{+}$/$1F\frac{7}{2}3^{+}$ and
$1D\frac{3}{2}2^{-}$/$1D\frac{5}{2}2^{-}$ states are concerned, they
are the mixings of the $^{3}F_{3}$/$^{1}F_{3}$ and
$^{3}D_{2}$/$^{1}D_{2}$ states respectively, and the mixture angle
can be determined by formula (10).

In order to distinguish $L$ from formula (8), we choose $l$ as the
orbital angular momentum of the $D$ mesons in the following three
formulas (12-14). If the initial states $A$($l^{P}$) are the
mixings, the partial wave amplitude can be deduced as

\begin{equation}
\begin{pmatrix} \mathcal{M}^{JL}_{|l-\frac{1}{2},l^{P}\rangle \rightarrow BC} &  \\ \mathcal{M}^{JL}_{|l+\frac{1}{2},l^{P}\rangle\rightarrow BC}  \end{pmatrix} \\=\begin{pmatrix} \cos\theta &  -\sin\theta  \\ \sin\theta & \cos\theta   \end{pmatrix} \\\begin{pmatrix} \mathcal{M}^{JL}_{|^{3}l_{l}\rangle\rightarrow BC} &  \\ \mathcal{M}^{JL}_{|^{1}l_{l}\rangle\rightarrow BC}  \end{pmatrix} \, , \\
\end{equation}
in the case of the mixings of the final states $B$($l^{'P^{'}}$)

\begin{equation}
\begin{pmatrix} \mathcal{M}^{JL}_{A\rightarrow |l^{'}-\frac{1}{2},l^{'P^{'}}\rangle C} &  \\ \mathcal{M}^{JL}_{A\rightarrow |l^{'}+\frac{1}{2},l^{'P^{'}}\rangle C}  \end{pmatrix} \\=\begin{pmatrix} \cos\theta^{'} &  -\sin\theta^{'}  \\ \sin\theta^{'} & \cos\theta^{'}   \end{pmatrix} \\\begin{pmatrix} \mathcal{M}^{JL}_{A\rightarrow |^{3}l^{'}_{l^{'}}\rangle C} &  \\ \mathcal{M}^{JL}_{A\rightarrow |^{1}l^{'}_{l^{'}}\rangle C}  \end{pmatrix}\, . \\
\end{equation}
When the initial and the final states ($A$ and $B$) are both the mixings, we can get the similar relation
\begin{equation}
\begin{pmatrix} \mathcal{M}^{JL}_{|l-\frac{1}{2},l^{P}\rangle\rightarrow |l^{'}-\frac{1}{2},l^{'P^{'}}\rangle C} &  \\ \mathcal{M}^{JL}_{|l-\frac{1}{2},l^{P}\rangle\rightarrow |l^{'}+\frac{1}{2},l^{'P^{'}}\rangle C}  & \\ \mathcal{M}^{JL}_{|l+\frac{1}{2},l^{P}\rangle\rightarrow |l^{'}-\frac{1}{2},l^{'P^{'}}\rangle C} & \\ \mathcal{M}^{JL}_{|l+\frac{1}{2},l^{P}\rangle\rightarrow |l^{'}+\frac{1}{2},l^{'P^{'}}\rangle C}\end{pmatrix} \\=\begin{pmatrix} \cos\theta\cos\theta^{'} &  -\sin\theta\cos\theta^{'} & -\cos\theta\sin\theta^{'} & \sin\theta\sin\theta^{'} \\ \cos\theta\sin\theta^{'} & -\sin\theta\sin\theta^{'} & \cos\theta\cos\theta^{'} &  -\sin\theta\cos\theta^{'} \\ \sin\theta\cos\theta^{'} & \cos\theta\cos\theta^{'} & -\sin\theta\sin\theta^{'} & -\cos\theta\sin\theta^{'} \\ \sin\theta\sin\theta^{'} & \cos\theta\sin\theta^{'} &\sin\theta\cos\theta^{'} &\cos\theta\cos\theta^{'}\end{pmatrix} \\\begin{pmatrix} \mathcal{M}^{JL}_{|^{3}l_{l}\rangle\rightarrow |^{3}l^{'}_{l^{'}}\rangle C} &  \\ \mathcal{M}^{JL}_{|^{1}l_{l}\rangle\rightarrow |^{3}l^{'}_{l^{'}}\rangle C} &  \\ \mathcal{M}^{JL}_{|^{3}l_{l}\rangle\rightarrow |^{1}l^{'}_{l^{'}}\rangle C} &  \\ \mathcal{M}^{JL}_{|^{1}l_{l}\rangle \rightarrow |^{1}l^{'}_{l^{'}}\rangle C}  \end{pmatrix} \, ,\\
\end{equation}
where $\theta$ and $\theta^{'}$ are the mixtures of the initial and
final states, respectively. Thus the decay width can also be deduced
from the general relations of (12-14). For example, in the case of
the mixings of the initial states of $1^{+}$,

\begin{equation}
\begin{pmatrix} \mathcal{M}^{JL}_{|\frac{1}{2},1^{+}\rangle\rightarrow BC} &  \\ \mathcal{M}^{JL}_{|\frac{3}{2},1^{+}\rangle\rightarrow BC}  \end{pmatrix} \\=\begin{pmatrix} \cos\theta &  -\sin\theta  \\ \sin\theta & \cos\theta   \end{pmatrix} \\\begin{pmatrix} \mathcal{M}^{JL}_{|^{3}P_{1}\rangle\rightarrow BC} &  \\ \mathcal{M}^{JL}_{|^{1}P_{1}\rangle\rightarrow BC}  \end{pmatrix} \, , \\
\end{equation}
and the decay width can be expressed as

\begin{equation}
\begin{split}
& \Gamma(|\frac{1}{2},1^{+}\rangle\rightarrow BC)=\frac{\pi}{4}\frac{|\vec P|}{M_{A}^{2}}\sum_{JL}|\cos\theta\mathcal{M}^{JL}_{|^{3}P_{1}\rangle\rightarrow BC}-\sin\theta\mathcal{M}^{JL}_{|^{1}P_{1}\rangle\rightarrow BC}|^{2} \, ,\\
& \Gamma(|\frac{3}{2},1^{+}\rangle\rightarrow BC)=\frac{\pi}{4}\frac{|\vec P|}{M_{A}^{2}}\sum_{JL}|\sin\theta\mathcal{M}^{JL}_{|^{3}P_{1}\rangle\rightarrow BC}+\cos\theta\mathcal{M}^{JL}_{|^{1}P_{1}\rangle\rightarrow BC}|^{2} \, . \\
\end{split}
\end{equation}

\begin{large}
\textbf{3 Numerical Results}
\end{large}

\begin{table*}[htbp]
\begin{ruledtabular}\caption{The adopted masses of the mesons used in our calculation.}
\begin{tabular}{c c c c c c c c c c c c c c c c c c}
States & \  $M_{\pi^{+}}$  & \ $M_{\pi^{0}}$  & \ $M_{K^{+}}$ &\ $M_{K^{*}}$ & \ $M_{\eta}$ & \ $M_{\eta^{'}}$ & \ $M_{D^{+}}$ & \ $M_{D^{0}}$  & \ $M_{D^{*+}}$ & \ $M_{D^{*0}}$ \\
Mass(MeV) & \   139.57         &  \ 134.98     & \  493.68    &  \   891.66   & \  547.85   &  \  957.78       &  \ 1869.6     &  \  1864.83    &  \ 2010.25  &  \  2006.96  \\
\hline
States & \ $M_{D^{*+}_{s}}$  & \ $M_{D^{+}_{s}}$ & \ $M_{D(2400)}$  & \ $M_{D(2430)}$ & \ $M_{D(2420)}$ & \ $M_{D(2460)}$ & \ $M_{D_{s}(2317)}$ & \ $M_{\rho}$ & \ $M_{\omega}$ & \  \\
Mass(MeV) & \    2112.3         &  \ 1968.47        & \  2318         &  \   2427        & \  2421.3       &  \  2464.4      &  \ 2317.8           &  \  770   &  \ 782  &  \  \\
\end{tabular}
\end{ruledtabular}
\end{table*}

The parameters involved in the $^{3}P_{0}$ decay  model include the
light quark pair ($q\overline{q}$) creation strength $\gamma$, the
SHO wave function scale parameter $R$, and the masses of the mesons
and the constituent quarks. According to Ref.~\cite{Blun}, we adopt
the SHO wave functions  with the common oscillator  parameter  $R$
whose value is chosen to be $2.5$ GeV$^{-1}$. Correspondingly, the
value of the $\gamma$  is chosen to be $6.25$ for the creation of
the $u/d$ quark~\cite{Blun,Dml10}. As for the strange quark pair
($s\overline{s}$), its creation strength can be related by
$\gamma_{s\overline{s}}=\gamma/\sqrt{3}$~\cite{Yao77}. The adopted
masses of the mesons are listed in TABLE IV, and $m_{u} = m_{d} =
0.22$ GeV, $m_{s} = 0.419$ GeV and $m_{c} = 1.65$ GeV.

\begin{table*}[htbp]
\begin{ruledtabular}\caption{The strong decay widths of the newly observed charmed mesons $D_{J}(2580)$, $D_{J}^{*}(2650)$, $D_{J}(2740)$ and $D_{J}^{*}(2760)$ with possible assignments. If the corresponding decay channel is forbidden, we mark it by "-". All values in units of MeV.}
\begin{tabular}{c c c c c c c c c c c c c c c c c c}
   & \ $D_{J}(2580)$  & \ $D^{*}_{J}(2650)$  & \ $D^{*}_{J}(2760)$ & \ $D_{J}(2740)$ & \ $D_{J}(2740)$ & \ $D^{*}_{J}(2760)$ &   \\
   & \ $2S\frac{1}{2}0^{-}$  & \ $2S\frac{1}{2}1^{-}$  & \ $1D\frac{3}{2}1^{-}$ & \ $1D\frac{3}{2}2^{-}$ & \ $1D\frac{5}{2}2^{-}$ & \ $1D\frac{5}{2}3^{-}$ &   \\
\hline
$D^{*+}\pi^{-}$       &  \  49.80      & \  34.72       &  \   16.46   & \  17.04     &  \  48.63    &  \ 10.35    \\
$D_{S}^{*+}K^{-}$     &  \    -    & \    2.02     &  \  2.86    & \  0.38     &  \  6.95    &  \  0.09   \\
$D^{*0}\pi^{0}$       &  \  25.00      & \  17.32       &  \  8.19    & \  8.74     &  \  24.24    &  \ 5.32    \\
$D^{*0}\eta$          &  \  0.81      & \   3.26      &  \  2.80    & \   0.59    &  \  7.41    &  \  0.24   \\
$D^{0}(2400)\pi^{0}$  &  \  0.35      & \    -     &  \   -   & \   0.06    &  \  0.00028    &  \  -   \\
$D^{0}(2460)\pi^{0}$  &  \   -     & \   0.024      &  \  0.23    & \  0.97     &  \  23.26    &  \  0.17   \\
$D(2420)\pi^{0}$      &  \   -     & \   0.12      &  \  23.92    & \  0.26     &  \  0.089    &  \  0.024   \\
$D(2427)\pi^{0}$      &  \   -     & \   0.30      &  \  3.10    & \   13.48    &  \  0.0057    &  \ 0.065    \\
$D^{+}\pi^{-}$        &  \   -     & \   13.57      &  \ 25.15     & \   -    &  \   -   &  \  17.02   \\
$D_{S}^{+}K^{-}$      &  \   -     & \   6.56      &  \  10.34    & \   -    &  \  -    &  \   0.70  \\
$D^{0}\pi^{0}$        &  \   -     & \   6.65      &  \  12.35    & \  -     &  \   -   &  \   8.73  \\
$D^{0}\eta$           &  \   -     & \   3.53      &  \  6.78    & \    -   &  \  -    &  \  0.99   \\
$D^{+}\rho$           &  \   -     & \   1.59      &  \   37.00   & \  36.02     &  \ 0.50     &  \  26.44   \\
$D_{S}^{+}K^{*}$      &  \   -     & \    -     &  \  -    & \   -    &  \   -   &  \  -   \\
$D^{0}\rho$           &  \   -     &  \  1.37      & \   19.10    & \  18.90     & \  0.29     &  \  13.72   \\
$D^{0}\omega$         &  \   -     & \   0.14      &  \  17.80    & \   16.90    &  \  0.20    &  \ 12.62    \\
Total width           &  \  75.96      & \  91.18       &  \  186.06    & \  113.34     &  \   111.56   &  \  96.49   \\
\end{tabular}
\end{ruledtabular}
\end{table*}

The numerical values of the widths of the strong decays of the
charmed mesons $D_{J} (2580)$, $D_{J}^{*} (2650)$, $D_{J} (2740)$,
$D_{J}^{*} (2760)$, $D_{J} (3000)$, $D_{J}^{*} (3000)$ observed by
the LHCb collaboration are presented in TABLE V-VII. The measurements
of the LHCb collaboration favor the assignment ($D_{J} (2580)$,
$D_{J}^{*} (2650)$) = $(0^{-}, 1^{-})_{\frac{1}{2}}$ with n = 2.
They also favor the following two possible assignments
\begin{equation*}
(D_{J}^{*} (2760),D_{J}(2740)) = (1^{-}, 2^{-})_{\frac{3}{2}}\quad  with \quad n = 1, L = 2 \, ,
\end{equation*}
\begin{equation*}
(D_{J} (2740),D_{J}^{*}(2760)) = (2^{-}, 3^{-})_{\frac{5}{2}}\quad  with \quad n = 1, L = 2 \, .
\end{equation*}
The partial and total decay widths in the above assignments are
listed in TABLE V. Comparing with the experimental data of the  LHCb
and BaBar collaborations, our results are of the same order in
magnitude. We can see that, except for the $D_{J} (2580)$, the
predicted total widths of the $D_{J} (2740)$ and $D_{J}^{*}(2760)$
are somewhat bigger than the experimental values, and the width of
the $D_{J}^{*} (2650)$ is roughly  in agreement with the total width
measured by the BaBar collaboration. In addition, the
$1D\frac{5}{2}3^{-}$ state  may be the optimal assignment of the
$D_{J}^{*} (2760)$ since the corresponding total width is close to
the experimental value  of the LHCb collaboration. However, the LHCb
collaboration identify  the $D_{J}^{*} (2760)$ as the
$1D\frac{3}{2}1^{-}$ state, which is incompatible with our results.
From TABLE V, we can see that, if the $D_{J}^{*}(2760)$ is the
$1D\frac{5}{2}3^{-}$ state, the main decay channels are $D^{+}\rho$,
$D^{+}\pi^{-}$, $D^{0}\rho$ and $D^{0}\omega$. The decay behavior of
the $1D\frac{3}{2}1^{-}$ state is very similar to that of the
$1D\frac{5}{2}3^{-}$ state except for the decay channel
$D(2420)\pi^{0}$. This difference can be used to further identify
the assignment of the $D_{J}^{*} (2760)$ in the future. Furthermore,
we tentatively identify the $D_{J}(2740)$ as the $1D$ state with
$J^{P}=2^{-}$, and we can see that the total widths of the
$1D\frac{3}{2}2^{-}$ and $1D\frac{5}{2}2^{-}$ states are of the same
order. However, the decay behaviors of these two states are
different from each other. The main decay modes of the
$1D\frac{3}{2}2^{-}$ state are $D^{+}\rho$, $D^{*+}\pi^{-}$,
$D^{0}\rho$, $D^{0}\omega$ and $D(2427)\pi^{0}$, while
 the $1D\frac{5}{2}2^{-}$ state  mainly decays  into
$D^{*+}\pi^{-}$,$D^{*0}\pi^{0}$ and $D^{0}(2460)\pi^{0}$.

\begin{table}[htbp]
\begin{ruledtabular}\caption{The strong decay widths of the newly observed charmed meson $D^{*}_{J}(3000)$ with possible assignments. If the corresponding decay channel is forbidden, we mark it by "-". All values in units of MeV.}
\begin{tabular}{c c c c c c c c c c c c c c c c c c}
       & \ $1F\frac{5}{2}2^{+}$  & \ $1F\frac{7}{2}4^{+}$  & \ $2P\frac{1}{2}0^{+}$ & \ $2P\frac{3}{2}2^{+}$ & \ $3S\frac{1}{2}1^{-}$   \\
\hline
$D^{*+}\pi^{-}$       &  \  10.46      & \  9.36       &  \   -   & \  11.91     &  \  3.19        \\
$D_{S}^{*+}K^{-}$     &  \  1.80    & \    0.27     &  \  -    & \  6.38     &  \  1.50      \\
$D^{*0}\pi^{0}$       &  \  5.21      & \  4.78       &  \  -    & \  5.86     &  \  1.66     \\
$D^{*0}\eta$          &  \  1.87      & \   0.54      &  \  -    & \   3.16    &  \  0       \\
$D^{*0}\eta^{'}$          &  \  0.10      & \   -      &  \  -    & \   0.25    &  \  1.84       \\
$D^{+}\pi^{-}$        &  \  12.61      & \   14.08      &  \ 23.94     & \  3.40     &  \  3.61       \\
$D_{S}^{+}K^{-}$      &  \  3.71      & \   0.81      &  \  2.85    & \  5.47     &  \   0.081     \\
$D^{0}\pi^{0}$        &  \  6.22      & \   7.19      &  \  11.97    & \   1.61    &  \   1.84     \\
$D^{0}\eta$           &  \   2.94     & \   1.21      &  \  4.26    & \   1.6    &  \    0.23     \\
$D^{0}\eta^{'}$           &  \ 4.39       & \   0.11      &  \  1.07    & \  4.31     &  \  1.69       \\
$D^{*+}\rho$           &  \   6.74     & \   28.23      &  \   62.01  & \  25.22     &  \ 18.01       \\
$D^{*+}_{S}K^{*}$      &  \   0.0057     & \   0.032      &  \  3.06    & \   12.21    &  \  0.57      \\
$D^{*0}\rho$           &  \   3.49     &  \  14.45      & \   31.60    & \  12.78     & \  8.73        \\
$D^{*0}\omega$         &  \   3.16     & \   13.46      &  \  29.91    & \   12.38    &  \  9.65        \\
$D^{+}\rho$           &  \    11.69    & \   2.78      &  \   -   & \  20.71    &  \ 0.14        \\
$D_{S}^{+}K^{*}$      &  \    0.61    & \    0.016     &  \   -   & \   2.29    &  \  3.53      \\
$D^{0}\rho$           &  \   5.90     &  \  1.46      & \   -    & \  10.37     & \  0.049        \\
$D^{0}\omega$         &  \   5.81     & \   1.33      &  \  -    & \   10.38    &  \  0.11        \\
$D(2420)\pi^{0}$      &  \   15.01     & \   0.04      &  \  26.20    & \  3.91     &  \  7.10       \\
$D(2420)\eta$      &  \   0.61     & \   $1.82\times10^{-7}$      &  \  1.37    & \  $1.62\times10^{-3}$     &  \  0.11       \\
$D(2427)\pi^{0}$      &  \  2.06      & \   0.41      &  \  6.69    & \   1.95    &  \  0.91        \\
$D(2427)\eta$      &  \   0.04     & \   $1.56\times10^{-4}$      &  \  0.35    & \  0.13     &  \  0.77       \\
$D(2400)\pi^{0}$  &  \  -      & \    -     &  \    -  & \   -    &  \  -     \\
$D(2400)\eta$  &  \  -      & \   -      &  \   -   & \   -    &  \  -     \\
$D_{S}(2460)K^{-}$     &  \    3.07    & \    0.016     &  \  12.81    & \  1.03     &  \  0.98      \\
$D_{S}(2536)K^{-}$     &  \    1.54    & \    0.0081     &  \  6.40    & \  0.64     &  \  0.49      \\
$D^{+}(2460)\pi^{-}$  &  \  4.85      & \   1.14      &  \  -    & \  11.08     &  \  13.57       \\
$D^{0}(2460)\pi^{0}$  &  \  2.45      & \   0.59      &  \  -    & \  5.59     &  \  6.87       \\
$D^{0}(2460)\eta$  &  \  0.09      & \   -      &  \  -    & \  0.05     &  \  0.008       \\
$D_{S}^{+}(2317)K^{-}$  &  \   -     & \   -      &  \  -    & \  -     &  \  -       \\
Total width           &  \  116.43      & \  102.31       &  \  224.49    & \  174.53     &  \   87.22      \\
\end{tabular}
\end{ruledtabular}
\end{table}

As discussed at the end of Section 1, the $D^{*}_{J}(3000)$ is a
natural parity state. Thus, we study its decay behavior with the
$1F\frac{5}{2}2^{+}$, $1F\frac{7}{2}4^{+}$, $2P\frac{1}{2}0^{+}$,
$2P\frac{3}{2}2^{+}$ and $3S\frac{1}{2}1^{-}$ assignments. We can
see from TABLE VI that the $D^{*}_{J}(3000)$ is most likely to be
the $1F\frac{5}{2}2^{+}$ state or  $1F\frac{7}{2}4^{+}$ state, since
the total widths are in good agreement   with the experimental data.
However, these two assignments  lead to different decay modes, which
can be used to further identify its quantum numbers. If the
$D^{*}_{J}(3000)$ is the $1F\frac{5}{2}2^{+}$ state, the
$D^{*+}\pi^{-}$, $D^{+}\pi^{-}$, $D^{+}\rho$ and $D(2420)\pi^{0}$
are the main decay modes, on the other hand, if  the
$D^{*}_{J}(3000)$ is the $1F\frac{7}{2}4^{+}$ state,  the
$D^{*+}\rho$, $D^{*0}\rho$, $D^{*0}\omega$, $D^{+}\pi^{-}$ and
$D^{*+}\pi^{-}$ are the main decay modes. Our results  show that the
assignments of the  $2P\frac{1}{2}0^{+}$, $2P\frac{3}{2}2^{+}$ and
$3S\frac{1}{2}0^{-}$ states can be excluded since the corresponding
total widths are quite different from the
 experimental values. Nevertheless, these information  are
valuable in searching  for the partners of the $D^{*}_{J}(3000)$. In
Ref.~\cite{Sun13}, Y. Sun \emph{et al}. identify the
$D^{*}_{J}(3000)$ as the $2^{3}P_{0}$ state with the effective
oscillator parameter $R$. In their studies, it is proposed that the
main decay channels of the $2^{3}P_{0}$ state are $D^{*}\rho$,
$D(2420)\pi$, $D(2427)\pi$, $D\eta$, $D_{S}K$, and $D^{*}\omega$.

As for the $D_{J}(3000)$, the possible assignments are the
$3S\frac{1}{2}0^{-}$, $2P\frac{1}{2}1^{+}$, $2P\frac{3}{2}1^{+}$,
$1F\frac{5}{2}3^{+}$ and $1F\frac{7}{2}3^{+}$ states. In TABLE VII,
the partial and total decay widths of the $D_{J}(3000)$ in those
possible  assignments  are given. We can see  easily from the table
that both the widths of the $1F\frac{7}{2}3^{+}$ and
$2P\frac{1}{2}1^{+}$ states  are in good agreement   with the
experimental data. So the $D_{J}(3000)$ is most likely to be the
$1F\frac{7}{2}3^{+}$ or $2P\frac{1}{2}1^{+}$ state. If the
$D_{J}(3000)$ is the $1F\frac{7}{2}3^{+}$ state, it dominantly
decays into $D^{+}(2460)\pi^{-}$, $D^{0}(2460)\pi^{0}$,
$D^{*+}\pi^{-}$, $D^{*0}\pi^{0}$ and $D^{*+}\rho$, on the other
hand,  if the $D_{J}(3000)$ is the $2P\frac{1}{2}1^{+}$ state, it
dominantly decays into $D^{*+}\rho$, $D^{*0}\rho$, $D^{*0}\omega$,
$D^{*+}\pi^{-}$, $D^{+}\rho$ and $D^{*0}\pi^{0}$. These conclusions
are consistent with the experimental observation\cite{Aaij}, where
the $D_{J}(3000)$ was firstly  observed in the $D^{*+}\pi^{-}$ decay
channel. As for the other three assignments $3S\frac{1}{2}0^{-}$,
$2P\frac{3}{2}1^{+}$ and $1F\frac{5}{2}3^{+}$, we can also see the
main decay modes from TABLE VII, which are valuable in searching for
these states   experimentally  in the future. In Ref.\cite{Sun13},
Y. Sun \emph{et al}. also suggest  that the $2P1^{+}$ state is the
mostly probable   assignment of the  $D_{J}(3000)$, which is
compatible with our observation. However, the $1F3^{+}$ assignment
is excluded in their studies as the width deviates    from the
experimental value. The differences between the results of Y. Sun et
al and ours are mainly due to the influence of the input parameter
$R$. And we will give a short discussion about the  dependence on
the $R$ at the end of this section.
\begin{table*}[htbp]
\begin{ruledtabular}\caption{The strong decay widths of the newly observed charmed meson $D_{J}(3000)$ with possible assignments. If the corresponding decay channel is forbidden, we mark it by "-". All values in units of MeV.}
\begin{tabular}{c c c c c c c c c c c c c c c c c c}
       & \ $1F\frac{5}{2}3^{+}$  & \ $1F\frac{7}{2}3^{+}$  & \ $2P\frac{1}{2}1^{+}$ & \ $2P\frac{3}{2}1^{+}$ & \ $3S\frac{1}{2}0^{-}$   \\
\hline
$D^{*+}\pi^{-}$       &  \  17.02      & \  25.95       &  \   20.32   & \  23.62     &  \  4.78        \\
$D_{S}^{*+}K^{-}$     &  \  0.57    & \    4.40     &  \  9.45    & \  1.22     &  \  2.25      \\
$D^{*0}\pi^{0}$       &  \  8.69      & \  12.93       &  \  10.03    & \  11.85     &  \  2.47     \\
$D^{*0}\eta$          &  \  1.05      & \   4.59      &  \  4.92    & \   2.48    &  \  0       \\
$D^{*0}\eta^{'}$          &  \  0.0057      & \   0.25      &  \  2.71    & \   18.72    &  \  2.75       \\
$D^{+}\pi^{-}$        &  \   -    & \    -     &  \  -   & \   -    &  \    -     \\
$D_{S}^{+}K^{-}$      &  \    -    & \   -      &  \  -    & \ -     &  \   -     \\
$D^{0}\pi^{0}$        &  \   -    & \    -     &  \    -  & \   -    &  \    -    \\
$D^{0}\eta$           &  \   -     & \    -     &  \   -   & \   -    &  \    -     \\
$D^{0}\eta^{'}$           &  \   -     & \   -     &  \  -    & \   -    &  \  -     \\
$D^{*+}\rho$           &  \   12.16     & \   12.79      &  \   41.34  & \  36.26     &  \ 15.44       \\
$D^{*+}_{S}K^{*}$      &  \   0.0081     & \   0.016      &  \  2.05    & \   4.08    &  \  0.49      \\
$D^{*0}\rho$           &  \   6.24    &  \  6.56      & \   21.07    & \  18.46     & \  7.48        \\
$D^{*0}\omega$         &  \   5.77     & \   6.07      &  \  19.93    & \   17.53    &  \  8.26        \\
$D^{+}\rho$           &  \    29.22    & \   5.01      &  \   10.59   & \  34.52    &  \ 0.21        \\
$D_{S}^{+}K^{*}$      &  \    1.50    & \    0.024     &  \   7.13   & \   3.82    &  \  5.31      \\
$D^{0}\rho$           &  \   14.74     &  \  2.63      & \   5.61    & \  17.27     & \  0.073        \\
$D^{0}\omega$         &  \   14.52     & \   2.39      &  \  4.99    & \   17.30    &  \  0.15        \\
$D(2420)\pi^{0}$      &  \   0.99     & \   0.40      &  \  0.0081    & \  0.024     &  \  -       \\
$D(2420)\eta$      &  \   $1.7\times10^{-3}$     & \    $6.3\times10^{-4}$     &  \ $3.0\times10^{-3}$   & \  $6.1\times10^{-3}$     &  \  -       \\
$D(2427)\pi^{0}$      &  \  0.0081      & \   $9.7\times10^{-3}$      &  \  $9.9\times10^{-3}$    & \   0.0081    &  \  -        \\
$D(2427)\eta$      &  \   $5.9\times10^{-4}$     & \   $1.8\times10^{-4}$      &  \  $1.5\times10^{-3}$    & \  $3.0\times10^{-3}$     &  \  -      \\
$D(2400)\pi^{0}$  &  \  0.32      & \    0.057     &  \    0.24  & \   0.17    &  \  0.51     \\
$D(2400)\eta$  &  \  0.011      & \   0.0027      &  \   0.27   & \   0.30    &  \  0.28     \\
$D_{S}(2460)K^{-}$     &  \    $1.9\times10^{-3}$    & \    $2.5\times10^{-3}$     &  \  0.0081    & \  0.024     &  \  -      \\
$D_{S}(2536)K^{-}$     &  \    $3.7\times10^{-3}$    & \    $4.9\times10^{-3}$     &  \  0.024    & \  0.049     &  \  -      \\
$D^{+}(2460)\pi^{-}$  &  \  0.99      & \   36.52      &  \  10.52    & \  56.21     &  \  27.15       \\
$D^{0}(2460)\pi^{0}$  &  \  0.50      & \   18.32      &  \  5.39    & \  28.05     &  \  13.74       \\
$D^{0}(2460)\eta$  &  \  0.019      & \   0.85      &  \  0.024    & \  0.56     &  \  0.013       \\
$D_{S}^{+}(2317)K^{-}$  &  \   0.049     & \   0.016      &  \  0.83    & \  0.52     &  \  0.27       \\
Total width           &  \  114.40      & \  187.49       &  \  177.46    & \  293.05     &  \   91.62      \\
\end{tabular}
\end{ruledtabular}
\end{table*}

From tables V-VII, we can also see that most of the ratios among
different decay channels are roughly  consistent with the results in
Ref.~\cite{Wang}. For example, the ratios
$\frac{\Gamma(D_{J}(2580)\rightarrow
D_{S}^{*+}K^{-})}{\Gamma(D_{J}(2580)\rightarrow D^{*+}\pi^{-})}$,
$\frac{\Gamma(D_{J}(2580)\rightarrow
D^{*0}\pi^{0})}{\Gamma(D_{J}(2580)\rightarrow D^{*+}\pi^{-})}$ and
$\frac{\Gamma(D_{J}(2580)\rightarrow
D^{*0}\eta)}{\Gamma(D_{J}(2580)\rightarrow D^{*+}\pi^{-})}$ from the
heavy meson effective theory are $0$, $0.51$ and $0.02$,
respectively~\cite{Wang}, which  are consistent with the present
results
  $0$, $0.52$, and $0.02$, respectively. In TABLE VIII, we also present
the experimental value of the ratio
$\frac{\Gamma(D_{2}^{*}(2460)\rightarrow
D^{+}\pi^{-})}{\Gamma(D_{2}^{*}(2460)\rightarrow D^{*+}\pi^{-})}$
for the well established meson $D_{2}^{*}(2460)$ from the
BaBar~\cite{Amo}, CLEO~\cite{Avery94,Avery90}, ARGUS~\cite{Albr89},
and ZEUS~\cite{Chek09} collaborations. The result based on the heavy
meson effective theory in the leading order approximation is also
listed in TABLE VIII. The present prediction $2.29$ based on the
$^{3}P_{0}$ decay model is in excellent agreement with the average
experimental value 2.35~\cite{Wang}. Furthermore, this result is
consistent with the prediction based on the heavy meson effective
theory. Finally, it needs to be noticed that the ratios among the
decay widths of different charmed mesons based on the $^{3}P_{0}$
decay model are roughly consistent with the experimental data. For
example, the predicted ratio $\frac{\Gamma _{D_{J}(2580)}}{\Gamma
_{D_{J}^{*}(2650)}}\approx0.83$  based on the $^{3}P_{0}$ decay
model, the corresponding experimental value is 1.27; the predicted
ratio $\frac{\Gamma _{D^{*}_{J}(2760)}}{\Gamma
_{D_{J}(2740)}}\approx0.86$,  the corresponding experimental value
is 1.0.
\begin{table*}[htbp]
\begin{ruledtabular}\caption{The experimental values and numerical result based on the leading order heavy meson effective theory (HMET) of the ratio $\frac{\Gamma(D_{2}^{*}(2460)\rightarrow D^{+}\pi^{-})}{\Gamma(D_{2}^{*}(2460)\rightarrow D^{*+}\pi^{-})}$ compared to our numerical result based on the  $^{3}P_{0}$ decay model}
\begin{tabular}{c c c c c c c c c c c c c c c c c c}
BaBar~\cite{Amo} & \  CLEO~\cite{Avery94}  & \ CLEO~\cite{Avery90}  & \ ARGUS~\cite{Albr89} &\ ZEUS~\cite{Chek09} & \ HMET~\cite{Wang} & \ This work  \\
\hline
$1.47 \pm 0.03 \pm 0.16$ & \   $2.2 \pm 0.7 \pm 0.6$         &  \ $2.3 \pm 0.8$     & \  $3.0 \pm 1.1 \pm 1.5$    &  \   $2.8\pm0.8^{+0.5}_{-0.6}$   & \  2.29  &  \  2.29     \\
\end{tabular}
\end{ruledtabular}
\end{table*}

\begin{figure}[h]
\begin{minipage}[t]{0.45\linewidth}
\centering
\includegraphics[height=5cm,width=7cm]{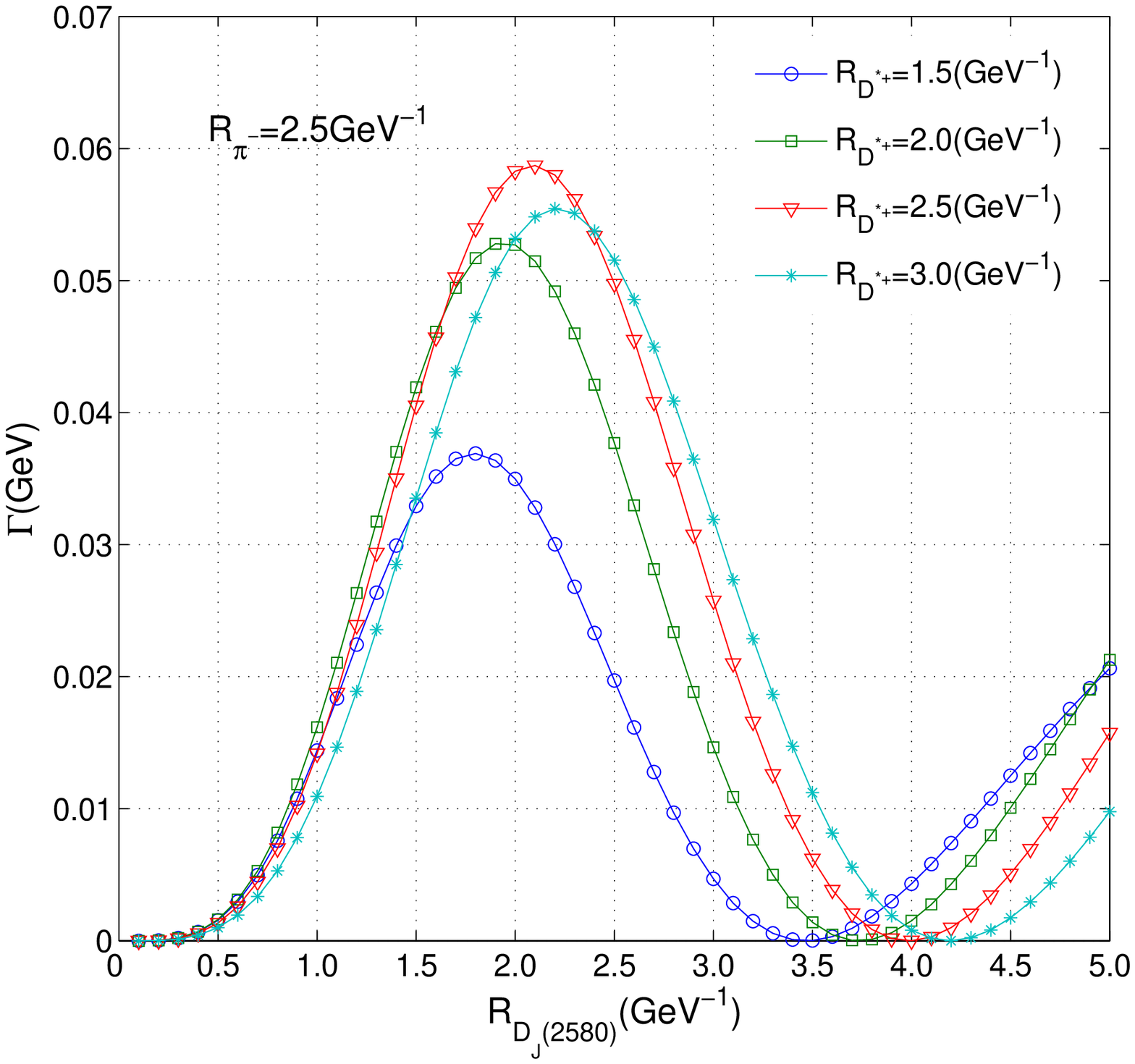}
\caption{The strong decay of $D_{J}(2580)\rightarrow D^{*+}\pi^{-}$
with $R_{\pi^{-}}=2.5$ GeV$^{-1}$.\label{your label}}
\end{minipage}
\hfill
\begin{minipage}[t]{0.45\linewidth}
\centering
\includegraphics[height=5cm,width=7cm]{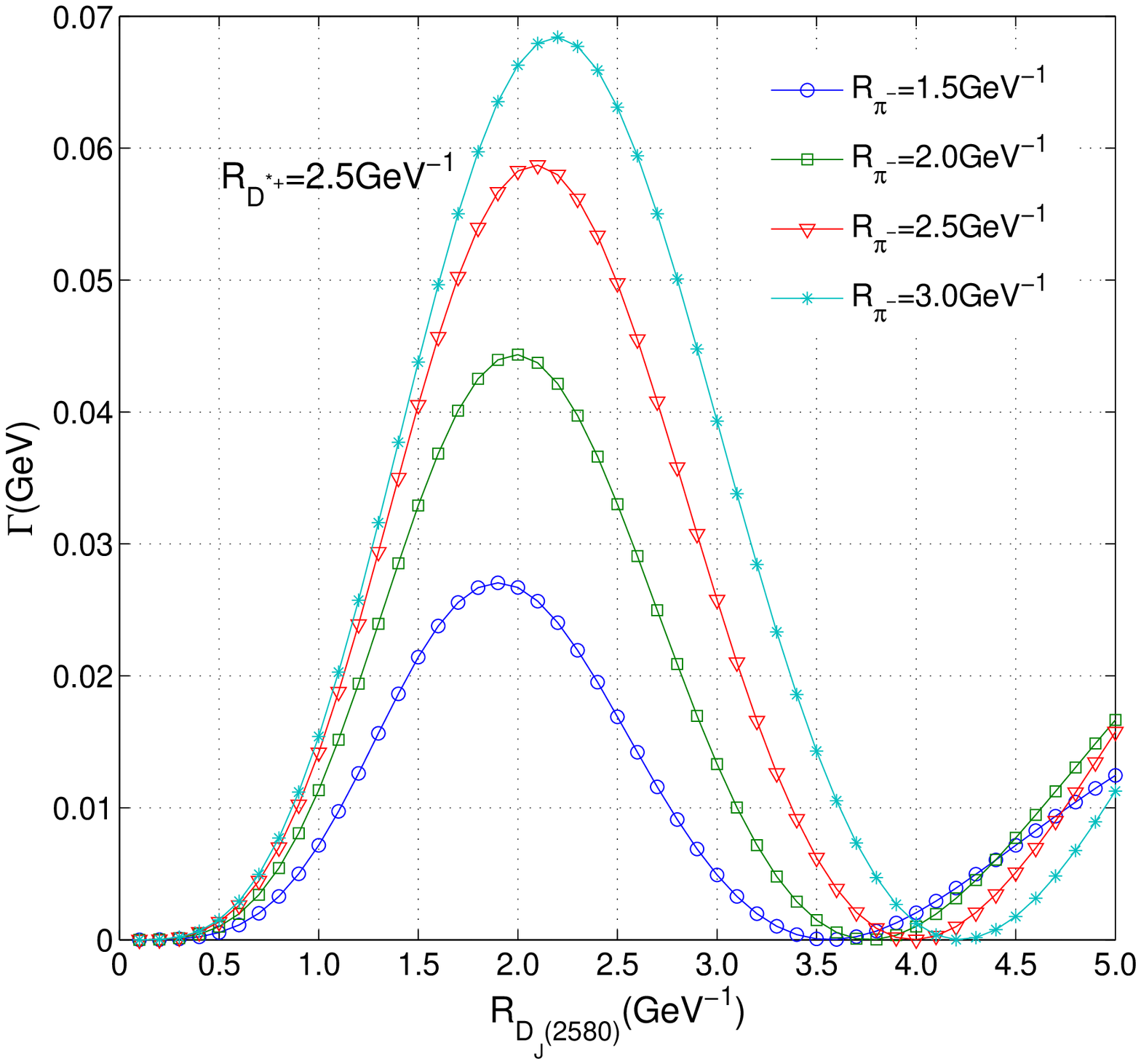}
\caption{The strong decay of $D_{J}(2580)\rightarrow D^{*+}\pi^{-}$
with $R_{D^{*+}}=2.5$ GeV$^{-1}$.\label{your label}}
\end{minipage}
\end{figure}

Now, let us take a short discussion about the uncertainties of the
results based on the $^{3}P_{0}$ decay model. Since this model is a
simplified model of a complicated theory, it is not surprising that
the prediction is not very accurate. Especially, the input parameter
$R$ has a significant influence on the shapes of the radial wave
functions, the spatial integral in equation (6) is sensitive to the
parameter $R$, therefore the decay width based on the $^{3}P_{0}$
decay model is sensitive to the  parameter $R$. We take the decay
$D_{J}(2580)\rightarrow D^{*+}\pi^{-}$ as an example, and plot the
decay width versus the input parameter $R$ in Figs. 2 and 3. From
these two figures, we can  see easily the dependence of the decay
width on the input parameter $R$. If the $R_{D^{*+}}$ and
$R_{\pi^{-}}$ are fixed to be $2.5$ GeV$^{-1}$, the decay width of
the $D_{J}(2580)$ changes several times with the value of the
$R_{D_{J}(2580)}$ changing from $1.5$ GeV$^{-1}$ to $3.0$
GeV$^{-1}$. Similarly, the decay width changes $2\sim3$ times, when
the $R_{D_{J}(2580)}$ and $R_{\pi^{-}}$(or the $R_{D_{J}(2580)}$ and
$R_{D^{*+}}$) are fixed to be $2.5$ GeV$^{-1}$ while the
$R_{D^{*+}}$(or $R_{\pi^{-}}$) changes. In Ref.~\cite{Blun}, H. G.
Blundel \emph{et al}  carry out a series of least squares fits of
the model predictions to the decay widths of 28 of the best known
meson decays. And the common oscillator parameter  $R$ with the
value of $2.5$ GeV$^{-1}$ is suggested to be the optimal
value~\cite{Blun}. As for the factor $\gamma$, it describes the
strength of quark-antiquark pair creation from the vacuum, which
also needs to be fitted according to experimental data, the fitted
value is 6.25~\cite{Blun,Dml10}.  Once the optimal values of the
$\gamma$ and $R$ are determined,
 the best predictions based on the ${}^3P_0$ decay model are expected
to be within a factor of 2. More detailed analysis about the
uncertainties of the results in the $^{3}P_{0}$ decay model can be
found in Ref.~\cite{Blun}.

\begin{large}
\textbf{4 Conclusion}
\end{large}

In this article, we   study   the properties of the charmed mesons
$D_{J}(2580)$, $D_{J}^{*}(2650)$, $D_{J}(2740)$, $D_{J}^{*}(2760)$,
$D_{J}(3000)$ and $D_{J}^{*}(3000)$ with the $^{3}P_{0}$ decay
model. Our results support the $1D\frac{5}{2}3^{-}$ assignment of
the  $D_{J}^{*}(2760)$, more experimental data are still needed to
identify it. Furthermore, both the mass spectra of the $D$ mesons
and the two-body decay behaviors indicate that the $D_{J}^{*}(3000)$
maybe the $1F\frac{5}{2}2^{+}$ state or $1F\frac{7}{2}4^{+}$ state,
since the widths in these two assignments are both  in good
agreement  with the experimental data. On the other hand, we
tentatively identify the $D_{J}(3000)$ as the $1F\frac{7}{2}3^{+}$
state and $2P\frac{1}{2}1^{+}$ state according to the decay widths.
It is noted that the $D_{J}^{*}(3000)$ and $D_{J}(3000)$ states are
strongly correlated to the background parameters as shown in
Ref.\cite{Aaij}.  Thus, more experimental data are still needed to
draw  a more clear conclusion on the existence of these two states.
In studying the $D_{J}(2580)$, $D_{J}^{*}(2650)$, $D_{J}(2740)$,
$D_{J}^{*}(2760)$, $D_{J}(3000)$ and $D_{J}^{*}(3000)$, we have also
obtained their  partial decay widths in different channels in the
assignments $2P\frac{1}{2}0^{+}$, $2P\frac{3}{2}2^{+}$,
$3S\frac{1}{2}1^{-}$, $3S\frac{1}{2}0^{-}$, etc, which can be used
to confirm or reject the assignments of the newly observed charmed
mesons in the future.

\begin{large}
\textbf{Acknowledgment}
\end{large}

This work is supported by National Natural Science Foundation of China, Grant Number 11375063, Natural Science Foundation of Hebei province, Grant Number A2014502017, and the Fundamental Research Funds for the Central Universities, Grant Number 13QN59.

\end{document}